\documentclass[aps,prb,citeautoscript]{revtex4}

\usepackage[pdftex]{graphicx}
\usepackage{amssymb,amsfonts,amsmath}
\usepackage{graphicx,amsmath}
\setcitestyle{super}
\renewcommand{\figurename}{Figure}

\begin{document}

\title{Room temperature magneto-optic effect in silicon light-emitting diodes}

\author{F. Chiodi$^{(a)}$, S.L. Bayliss$^{(b,c)}$, L. Barast$^{(a,b)}$, D. D\'ebarre$^{(b)}$, H. Bouchiat$^{(b)}$, R.H. Friend$^{(c)}$ and A.D. Chepelianskii$^{(b)}$\footnote{To whom correspondence should be addressed: alexei.chepelianskii@u-psud.fr}}

\affiliation{$(a)$ Centre de Nanosciences et de Nanotechnologies, CNRS, Univ. Paris-Sud, Universit\'e Paris-Saclay, C2N-Orsay, 91405 Orsay cedex, France}
\affiliation{$(b)$ LPS, Univ. Paris-Saclay, Univ. Paris-Sud, CNRS, UMR 8502, F-91405, Orsay, France}
\affiliation{$(c)$ Cavendish Laboratory, University of Cambridge, J J Thomson Avenue, Cambridge CB3 OHE, UK}

\begin{abstract}
In weakly spin-orbit coupled materials, the spin-selective nature
of recombination can give rise to large magnetic-field effects, for
example on the electro-luminescence of molecular semiconductors. While silicon has weak spin-orbit coupling, observing spin-dependent recombination through magneto-electroluminescence is challenging: silicon's indirect band-gap causes an inefficient emission, and it is difficult to separate spin-dependent
phenomena from classical magneto-resistance effects. Here we overcome these challenges and measure magneto-electroluminescence in silicon light-emitting diodes fabricated via gas immersion laser doping. These devices allow us to achieve efficient emission while retaining a well-defined geometry thus suppressing classical magnetoresistance effects to a few percent.
We find that electroluminescence can be enhanced by up to 300\% near room temperature in a seven Tesla magnetic field, showing that the control of the spin degree of freedom can have a strong impact on the efficiency of silicon LEDs.
\end{abstract}

\maketitle

{\it Introduction}

Spintronic effects in systems with weak spin-orbit coupling have attracted
considerable attention due to their rich fundamental physics and potential
for device applications \cite{Dyakonov2008,Jansen2012,Forrest2006,Awschalom2007}. 
A class of these effects can be measured
optically \cite{Schirhagl2014,Coquillat,Lepine1972}, providing direct insight into phenomena such as spin-dependent recombination, where only the singlet state of an electron-hole pair can recombine radiatively back to the ground state. Since external magnetic fields can change the spin statistics and energy levels in the sample, magneto-electroluminescence (MEL) effects have been seen as the hallmark of spin-dependent recombination phenomena, and have given important insight into the role of spin in organic materials
used for light-emitting diodes (LEDs) \cite{Vardeny2012,Shinar,Wang2012}. 
These spintronic effects can then be harnesseed, to provide very senstive magnetic field sensors, sensible to external magentic fields of only a few mTesla comparable with the fluctuating hyperfine fields inside organic materials \cite{Vardeny2010,Bobbert,Wohlgenannt2012} or to engineer new light-emitting device architectures through reverse intersystem crossing \cite{Adachi2014}.

Like molecular semiconductors, silicon has weak spin-orbit coupling, but emission is much less efficient due to silicon's indirect band-gap, making analogous
magneto-optic studies challenging, and requiring careful engineering
to prepare efficient light-emitting diodes \cite{Homewook,Green,Saito2015}.
In addition, observing spin-dependent magneto-electroluminescence in silicon
requires that the magnetic field and device currents are parallel to effectively suppress classical magnetoresistance (MR) contributions which can  
enhance MR in silicon up to spectacular values even at room temperature \cite{Kobayashi,CaihuaWan,DezhengYang,Porter,Schoonus}. 

Here we address both of these challenges by developing a new fabrication method for
efficient silicon light-emitting diodes using an original doping technique,
gas immersion laser doping (GILD), and investigate spin-dependent recombination
in silicon LEDs (SiLEDs). The GILD process \cite{Boulmer,Bhaduri,APT,Chiodi}
allows us to reach doping levels well beyond the solubility threshold
which, as we describe below, gives rise to efficient emission, while
retaining the well-defined planar geometry necessary to align electric
and magnetic fields. Using our SiLEDs, we find that when classical
MR effects are suppressed, electroluminescence
can be substantially enhanced under a magnetic field near room temperature. 
We explain this phenomenon
using a model of spin-dependent recombination \cite{kaplan1978,merrifield,Bayliss2015} of electron-hole pairs and use our analysis to estimate the exchange energy of weakly bound excitons in silicon. Our experiments provide an optoelectronic
approach to probe the spin statistics of carriers in silicon - a material which is an
excellent candidate for scalable spin quantum computing \cite{Kane1998,Leo2004,Tyryshkin2012}. They also highlight the importance of controlling the spin degree of freedom for the efficiency of silicon light emitting devices.

{\it Results}

\begin{figure}
\centerline{\includegraphics[clip,width=10cm]{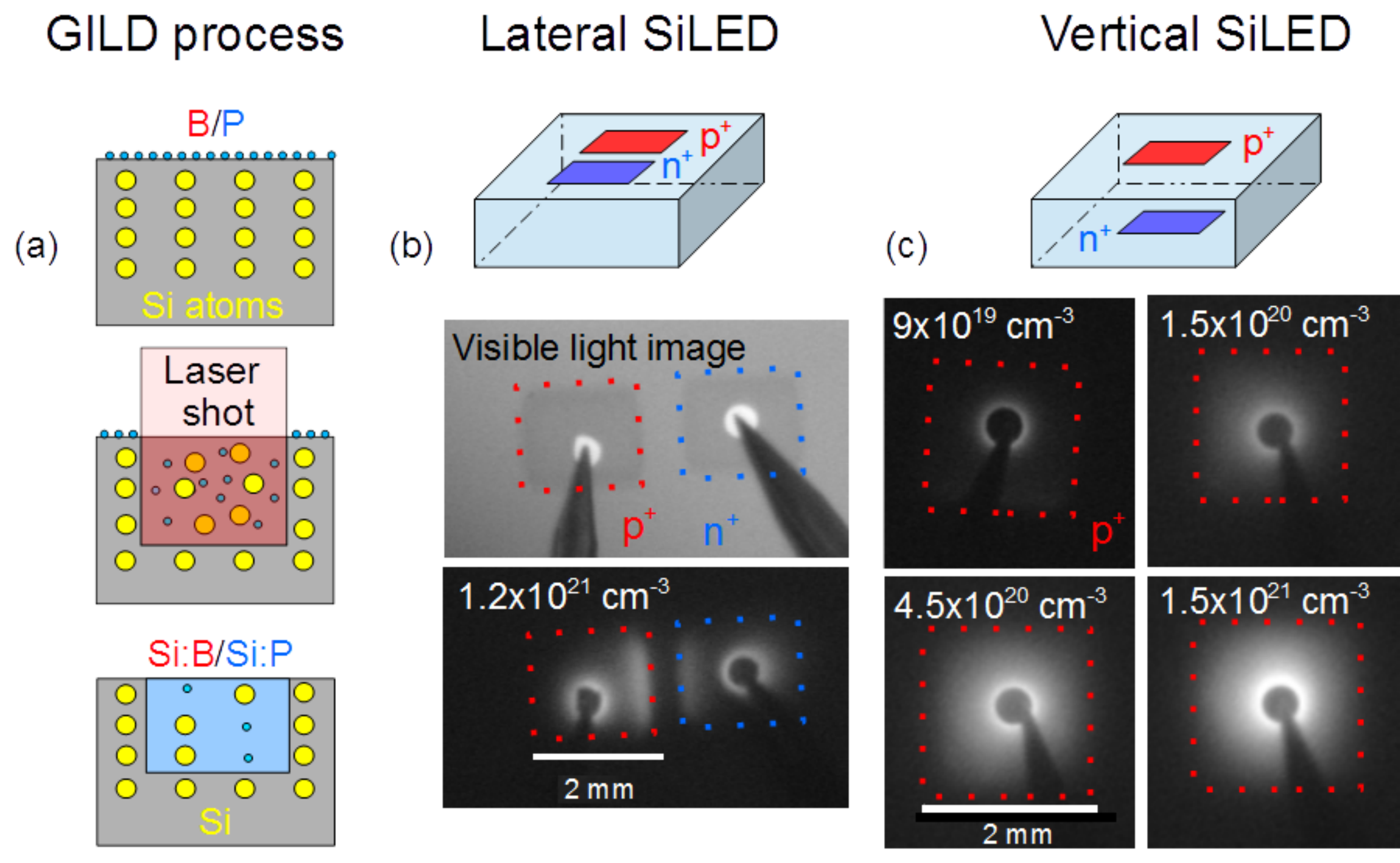}} \caption{
Silicon light-emitting diodes from gas immersion laser doping (GILD). (a) Schematics of the GILD doping process: chemisorbtion of the precursor gas (${\rm P Cl_3}/{\rm B Cl_3}$); laser melting of the substrate and dopant diffusion in the liquid phase; solidification and epitaxy of a Si:P or Si:B crystal.
(b) Schematic of lateral devices and infrared images of a Silicon light-emitting device (SiLEDs)  ($1.2\times10^{21}{\rm cm^{-3}}$) biased at 20 mA and at room temperature.
(c) Schematic of vertical devices and infrared images of SiLEDs biased at 20 mA at room temperature for different doping levels  ($9\times10^{19},1.5\times10^{20},4.5\times10^{20},1.5\times10^{21}{\rm cm^{-3}}$). 
}
\label{FigSetup} 
\end{figure}

{\bf Description of the system}: We start by describing the fabrication procedure of the GILD SiLEDs [Fig. 1-a]
and the physical mechanism behind their enhanced efficiency, before
discussing the MEL response of these devices.
The Si light-emitting diodes were prepared by doping two $2\times 2{\rm \, mm}^2$ spots with
opposite polarities p+/n+ on a n-Si [100] substrate of resistivity $45\;{\rm \Omega cm}$ and thickness $700\;{\rm \mu m}$ using
the GILD technique [Fig. 1-a]. A precursor
gas ${\rm P Cl_3}\, ({\rm B Cl_3})$ for n+ (p+) doping is injected into an ultra-high
vacuum chamber, where it saturates the chemisorbtion
sites on the Si surface.
The substrate is then melted by a pulsed excimer XeCl 308nm laser
with a 25 ns pulse duration. The dopants diffuse into the liquid Si
phase and are incorporated in the lattice as the liquid/solid interface
moves back to the surface at the end of the irradiation \cite{APT}. A Si:P/Si:B
crystal is thus created by fast liquid phase epitaxy above the underlying
Si substrate [Fig. 1-a]. The dose of active dopants is determined
exclusively by the number of laser shots while the doping depth can
be independently tuned by controlling the laser energy.
Due to the short pulse duration
and high recrystallization speed, high dopant
concentrations beyond the solubility limit ($10^{20}{\rm cm^{-3}}$ for Si:B) \cite{Murrell} can be reached without introducing defects. 
In our experiments, we varied the doping concentration in the range
$4.5\times10^{19}\;{\rm to}\;5\times10^{21}{\rm cm^{-3}}$ while keeping the doping depth constant at 80 nm (pulse energy $960\;{\rm mJ\;cm^{-2}}$). Ti(15nm)/Al(200nm) electrodes were deposited on top of the doped spots after BHF deoxidation.
\begin{figure}
\centerline{\includegraphics[clip,width=10cm]{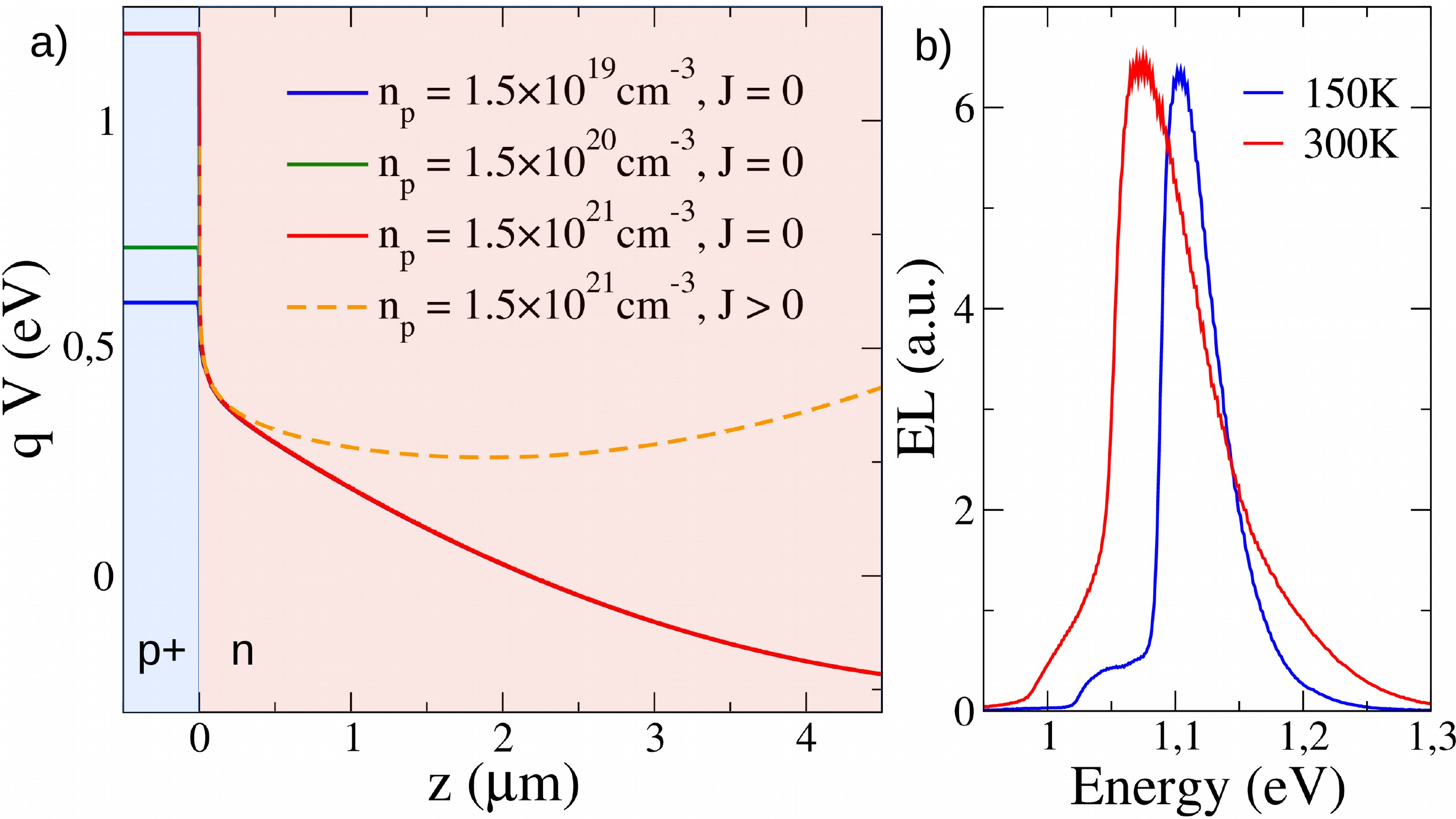}}
\caption{Origin of enhanced emission in gas immersion laser doping silicon light emitting devices. (a) Simulated electrostatic potential at the p+/n interface as a function of the vertical distance $z$ to the p+/n interface. 
Full curves show different p+ doping levels without external bias. The dashed curve shows the formation of a potential minimum under forward bias at a current density of $J=2\times 10^4 \;{\rm A m^{-2}}$ for the highest doping level. (b) Emission spectra from a lateral device with $1.2\times10^{21}{\rm cm^{-3}}$ doping.
 }
\label{FigEL} 
\end{figure}

{\bf Device electroluminescence}:  We investigated EL from GILD SiLEDs in two different device geometries where the p+ and n+ spots were placed laterally, without overlap, on the same Si surface (Fig. 1-b) and in a vertical geometry where the n+/p+ were prepared on top of each other on opposite sides of the Si wafer (Fig. 1-c). The alignment between top and bottom spots in the vertical geometry was achieved by looking through the silicon wafer with an infrared camera which allowed to see the bottom GILD spots. This camera also allowed to record characterization images of EL devices.
Images in the lateral geometry (Fig. 1-b) show that EL occurs
mainly at the heavily doped GILD spots on both n+ and p+ sides whereas
the undoped region in-between the two spots remains dark. This shows
that EL is enhanced near the n+ and p+ interfaces as compared to bulk
Si. This observation is supported by the strong improvement of
the device brightness as the GILD doping concentration increases (see Fig. 1-c). The external quantum efficiency (EQE) for our brightest devices is around $0.05\%$ which is comparable with the highest reported values for devices without anti-reflection treatement \cite{Homewook}.

The physical origin of the EL enhancement can be understood from the
electrostatic profile within the devices which we model using drift-diffusion
simulations of the p+/n interface accounting for the Fermi-statistics
in the highly doped regions \cite{Ansgar}. The simulated distributions of the electrostatic
potential $V(z)$ are shown on Fig. 2-a. In unbiased devices (see Supplementary Figure 1 for a more detailed discussion), the potential $V(z)$ inside the $n$ region ($z>0$) is almost independent
on the p+ doping concentration $n_{p}$; however a steep potential step forms at the p+/n interface, its height
increasing with doping, creating a barrier that electrons have to overcome to leave the device. 
For parameters corresponding to the brightest devices the
barrier is near an ${\rm eV}$ high and thermally activated transport is effectively prohibited.
The p+ region thus plays the role of an electron blocking layer while the n+ region will similarly act as a hole blocking layer. Such layers are known to enhance the efficiency of organic LEDS \cite{Ikai,Hagen}. When devices are biased the potential $V(z)$ remains constant in the p+ region and near
the p+/n interface as the applied potential will mainly
drop across the intrinsic weakly doped regions which have much larger
resistivity than the highly doped p+ region. A potential minimum therefore appears at sufficiently high forward
bias near the p+/n and n/n+ interfaces. 
In these regions, located at a vertical distance of $1-2\;{\rm \mu m}$ away from the interfaces,
the internal electric field vanishes favoring radiative recombination since a built-in electric field would otherwise drive electron-hole pairs apart.
Similarly the minority carriers (holes) predominantly recombine at the n+ interface as can be seen from the weaker EL observed from the n+ spot in lateral devices (see Fig.~1). Finally EL spectra in Fig.~2 (see also Supplementary Figure 2) are in very good agreement with previously reported lineshapes for bulk Si \cite{Tsybeskov,Jang,Lin} in agreement with our model predicting emission in the weakly doped Silicon a micrometer away from the p+/n and n+/n interfaces. 

\begin{figure}
\centerline{\includegraphics[clip,width=10cm]{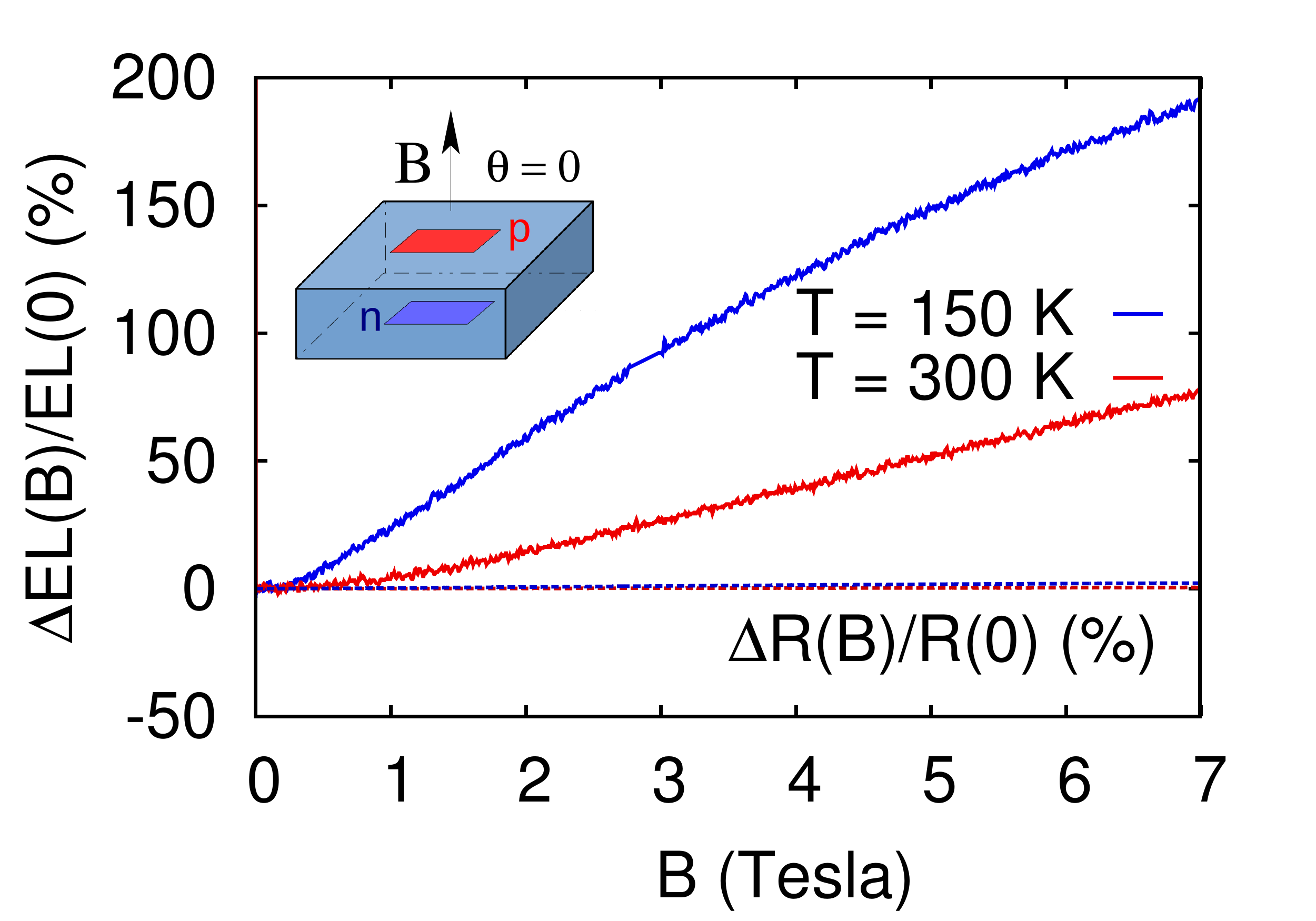}}
\caption{Magneto-electroluminescence in Silicon light emitting devices. Comparison between MEL and MR effects in a vertical SiLED under a perpendicular magnetic field (magnetic field tilt angle $\theta = 0$ relative to current lines) at $300 {\rm K}$ and $150 {\rm K}$ where $\Delta EL(B)=EL(B)-EL(0)$ and $\Delta R(B)=R(B)-R(0)$. The DC forward bias current was 10 mA. 
} 
\label{FigELI} 
\end{figure}

{\bf SiLED Magneto-electroluminescence}:  Having described the structure of our devices and the physical mechanism
behind their enhanced EL efficiency, we now use them to study the
dependence of the EL on magnetic field in a vertical SiLED device (doping $3\times10^{21}{\rm cm^{-3}}$); similar data was obtained on a device with $1.5\times10^{21}{\rm cm^{-3}}$ doping. For this experiment devices where mounted inside an optical access
magnet (Oxford instruments) and the EL was collected by a Ge photo-detector
outside the cryostat. The SiLEDs were DC biased and the input of the Ge photo-detector was chopped at $230\;{\rm Hz}$
to enhance sensitivity.

To study the role of spin in the MEL, we first applied a magnetic field $B$ perpendicular to the device surface, i.e. parallel to the
internal electric field. We obtained a vanishing classical
magetoresistance (MR),
as shown in Fig.~3 where we observe only a weak residual MR in
the 1\% range both at $300\;{\rm K}$ and $150\;{\rm K}$.
The accuracy of this MR cancellation can seem surprising given the lack of electrical insulation around the spots. However, close to the onset
voltage of the diode, the voltage drop mainly occurs accross the few micron wide depletion region between the p+ and n regions (the extent of the depletion
region is shown on Fig. Supp. 1) and the current lines cannot bend significantly considering the large spot size 2mm $\times$ 2mm.
Fig.~3 also plots the MEL response, which shows a drastically different behaviour. Compared to the MR signal, the EL exhibits a two orders of magnitude stronger dependence on the magnetic field, with $\Delta EL(B)/EL(0)=75\%$ at room temperature and an even higher $\Delta EL(B)/EL(0)=290\%$ value at $150\;{\rm K}$ (for a 5 mA current). 
The striking difference in magnitude of the MEL signal over the MR suggests that the
magnetic field is increasing radiative recombination. Since
the EL quantum efficiency is low in silicon a strong change in EL has little effect on
the total current, and hence the MR. 

\begin{figure}
\centerline{\includegraphics[clip,width=12cm]{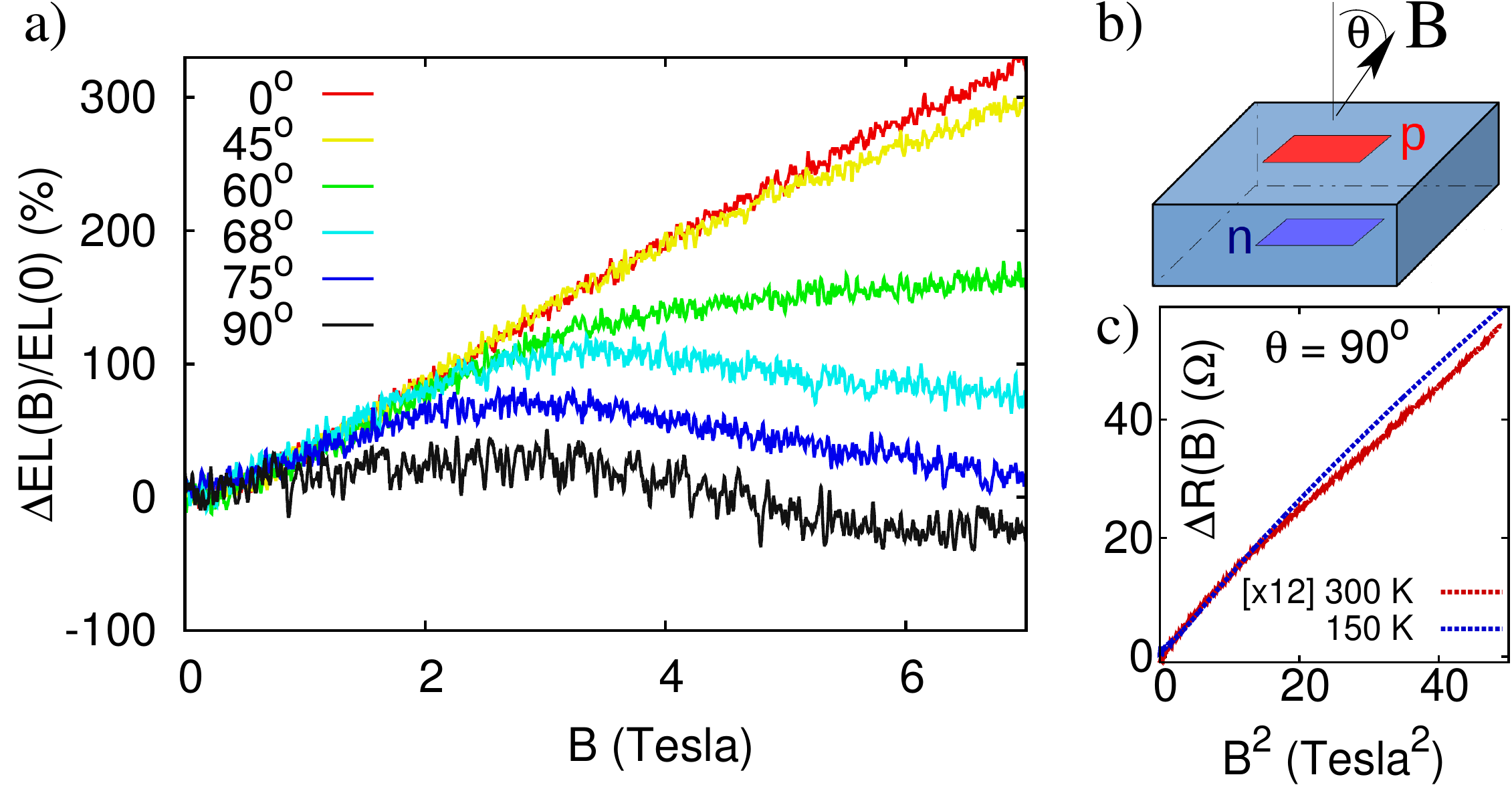}}
\caption{Angular dependence of magneto-electroluminescence in Silicon light emitting devices. Evolution of the magneto-electroluminescence as a function of the tilt angle $\theta$ between the magnetic field and the $5{\rm mA}$ DC current at 150 K (data shown in panel a, experiment geoetry is sketched in panel b). The panel c) shows the MR response measured at $\theta = 90{\rm ^{o}}$ which displays the classical $B^{2}$ dependence which contrasts with the MEL field dependence.}
\label{FigAngle} 
\end{figure}

To further investigate the origin of the strong MEL effect
in SiLEDs, we measured the MEL response as a function of the angle of the magnetic field relative to current
lines Fig.~\ref{FigAngle}. For moderate tilt angles $\theta$ below $45^{\circ}$ the
MEL signal is nearly independent on the tilt. At larger tilt angles
$\theta$ below $75^{\circ}$, the MEL is unchanged at low-magnetic fields
but decreases from the low tilt angle behaviour 
at high magnetic fields. 
 We attribute this decrease to the magneto-diode
effect \cite{Popovic,Stafeev,Cristo}: the in-plane component of the magnetic field bends electron and hole trajectories so that they cross a larger distance through the device and thus have a higher non radiative-recombination probability. A significant bending of current lines can occur in perpendicular electric and magnetic fields when the ratio $\mu_{xy}/\mu_{xx}$ between Hall and longitudinal mobility is large. This quantity is around $1$ at room temperature for $B = {\rm 7\,T}$ and around $5$ at 150K (using the mobility values for a $45 {\rm \, \Omega cm}$ electron doped Si: $\mu_{xx} = 1.4 \times 10^{3}\;{\rm cm^2 V^{-1} s^{-1}}$ at room temperature and $\mu_{xx} = 7 \times 10^{3}\;{\rm cm^2 V^{-1} s^{-1}}$ at 150 K). The negative magneto-diode contribution to MEL is further enhanced in lateral devices with longer current paths through the sample (see Supplementary Figure 4), with a negative MEL amplitude comparable with the ratio $\mu_{xy}/\mu_{xx}$. 

The full MEL response is thus a superposition of a positive MEL component which depends only weakly on the $B$-field direction, and a negative magneto-diode effect component which contributes more strongly at large tilt angles. A purely magnetodiode contribution would have the same $B$ dependence as the MR, which increases with the parallel magnetic field as $B^{2}$, as expected from the Drude law (see Fig.~4), in contrast with the observed MEL signal
which displays a linear dependence with $B$ except at the lowest fields.
Since the angle and magnetic field dependence of the positive MEL effect are very different from the MR, we infer that the MEL is not determined by transport scattering times and may be related to spin-degrees of freedom.

{\it Discussion} 

\begin{figure}
\centerline{\includegraphics[clip,width=12cm]{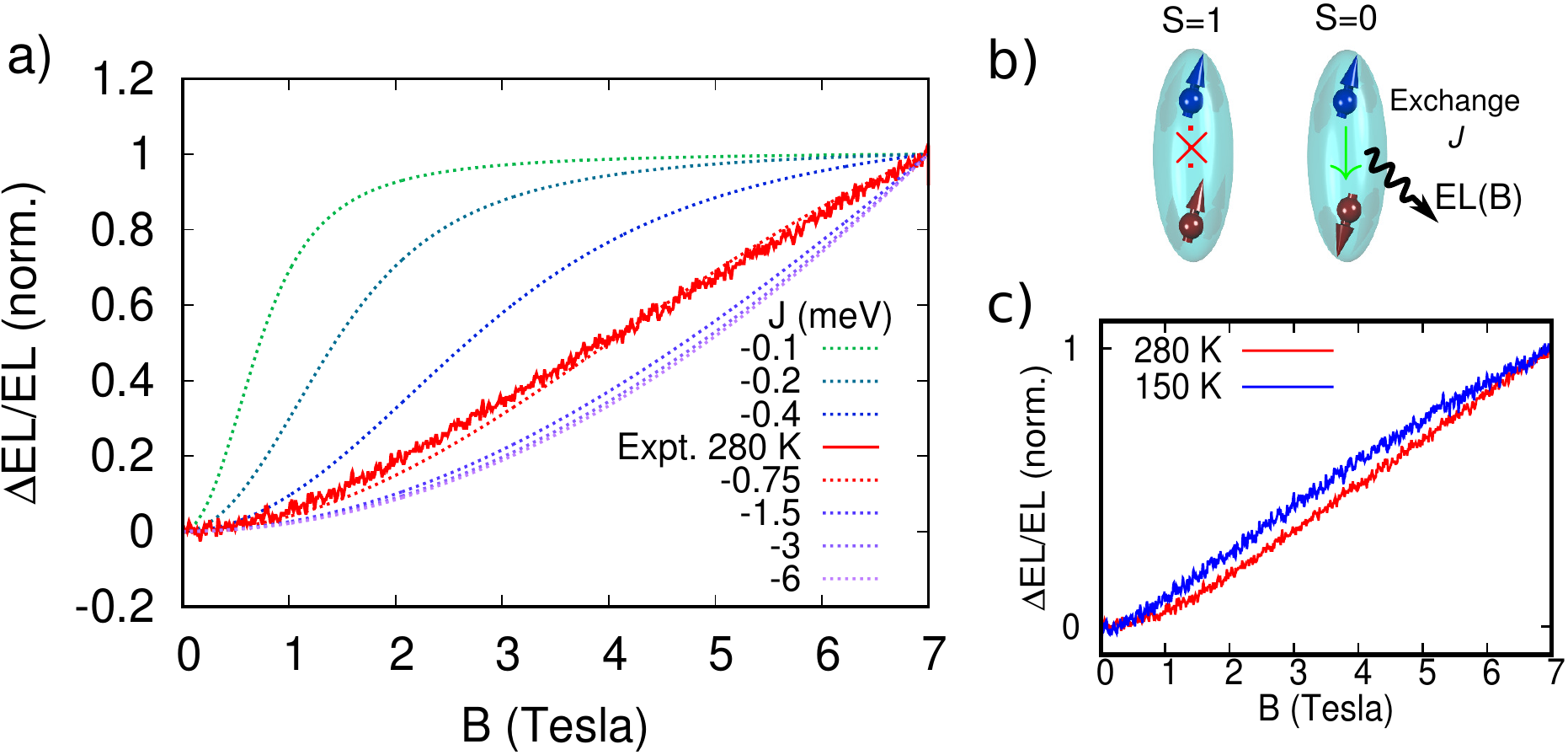}} 
\caption{Electron-hole spin-dependent recombination theory for the magneto-electroluminescence. Simulated magneto-electroluminescence using the model described in the text. Normalised simulations are shown in panel a) for varying electron-hole exchange energy $J$ alongside the experimental data at 300 K (a comparison between normalised 300K and 150K data is shown in panel c). The characteristic saturation field of the magneto-electroluminescence is determined by the electron-hole exchange energy. Fitting to the 300 K experimental data gives an exchange energy $J=-0.75$ meV. Panel b) illustrates the singlet and triplet spin pairings of a weakly bound electron/hole pair for which only the singlet state is emissive.}
\label{FigTheo} 
\end{figure}

The MEL effect observed in SiLEDs when the current and magnetic
field are parallel ($\theta = 0$) can be explained
as arising from the spin-dependent recombination of weakly bound electron-hole
pairs within the devices, in analogy with the models developed by
Kaplan et al. \cite{kaplan1978} and Merrifield \cite{merrifield}. The ability of an electron-hole pair to recombine radiatively is determined
by the overlap of the electron-hole pair wavefunction with the spin-zero
singlet ground state - radiative recombination will only be efficient for electron-hole pairs in an $S=0$ singlet
configuration, the recombination from an $S=1$ triplet electron-hole
pair being much less efficient (Fig.~\ref{FigTheo}-b). As the magnetic field is changed, the electron-hole
eigenstates - which are in general not pure singlet or triplet spin
states - are modified, altering their singlet overlap. As we explain
below, this change
in the electron-hole pair wavefunctions gives rise to a change in
EL, and can explain the lineshapes observed experimentally. Importantly,
this effect is independent of the direction of the magnetic field
and can therefore describe the angular-dependent MEL observations (Fig. 4). We note that
while spin-dependent free-carrier recombination has previously been
studied through circularly polarised emission \cite{li2010theory,sircar2014experimental,jonker},
here we invoke spin-dependent recombination of weakly bound exchange coupled electron-hole
pairs to explain our MEL effects.

We start by considering the kinetic equation for the population of transient electron-hole
pairs that are formed by brief collisions in the device recombination zone near room temperature.
The population $X_n$ of transient pairs with spin-eigenstates ${|n\rangle}$ is expected to follow the following rate equations:
\begin{align}
\dot{X}_{n}=G_{n}-\gamma_{s}\alpha_{n}X_{n}-\gamma X_{n}.
\end{align}

Here $G_{n}$ is the electron-hole pair generation rate. The second term describes the probability of (spin-dependent) radiative recombination during collision events, with a rate $\gamma_{s}$.
This term is proportional to the overlap of the electron-hole pair wavefunction with the spin-singlet state $\alpha_{n}=|\langle S|n\rangle|^{2}$, where $|S\rangle$ is the singlet state. The final (spin-independent) $\gamma$ term reflects both the probability of escape from the shallow potential well where radiative recombination occurs in our devices
and the rate of nonradiative relaxation.
Solving this in steady-state
for the total emission from the electon-hole pairs $EL=\sum_{n}\gamma_{s}\alpha_{n}X_{n}$
we find 
\begin{align}
EL\propto\sum_{n}\frac{G_{n}\alpha_{n}}{1+\epsilon\alpha_{n}},\label{eq:EL}
\end{align}
where $\epsilon=\gamma_{s}/\gamma$. This sum depends non-linearly
on the singlet projections $\{\alpha_{n}\}$ and so a magnetic-field
induced change in these can give rise to a change in emission. To
compute electroluminescence from Equation \ref{eq:EL}, we calculate the singlet projections by
diagonalising the following electron-hole pair spin-Hamiltonian 
\begin{align}
\hat{H}=\underbrace{g_{e}\mu_{B}\mathbf{B}\cdot\hat{\mathbf{S}}_{e}}_{electron}+\underbrace{g_{h}\mu_{B}\mathbf{B}\cdot\hat{\mathbf{S}}_{h}+\lambda\mathbf{\hat{L}}_{h}\cdot\mathbf{\hat{S}}_{h}}_{hole}+\underbrace{J\mathbf{\hat{S}}_{e}\cdot\hat{\mathbf{S}}_{h}}_{exchange},
\end{align}
where $g_{e},g_{h}$ are electron and hole $g$-factors, $\mu_{B}$
the Bohr magneton, $\mathbf{B}$ the applied field, $\mathbf{\hat{S}}_{e},\hat{\mathbf{S}}_{h}$
are the electron and hole spins, $\lambda$ is the spin-orbit parameter
for the hole with \textbf{$\hat{\mathbf{L}}_{h}$} the hole orbital
angular momentum, and $J$ is the electron-hole exchange coupling.
We note that the form of the effective spin-Hamiltonians are similar for the transient interacting electron hole pairs considered here and tightly bound excitons, the main difference being the amplitude of the exchange interaction $J$ compared to temperature with $J \ll k_B T$ for transient electon-hole pairs as opposed to bound excitons.

Fig.~5 shows the change in EL as a function of magnetic field calculated from Equation 2 for various values of the electron-hole exchange parameter $J$, using $\text{\ensuremath{\lambda}=-44}$ meV \cite{SiBand,Madelung}, $g_{e}=g_{h}=2$ \cite{Leo2009,Silvano2015,Silvano2015}, Boltzmann populations i.e., $G_{n}\propto e^{-E_{n}/k_{B}T}$ where $E_{n}$ is the electron-hole pair energy, $k_{B}$ is Boltzmann's constant, and $T=300$ K is the temperature. (This population distribution assumes that electron-hole pairs thermalise within their encounter time, but due to the high temperatures involved, similar results are obtained in the fully unpolarized limit where the generation rates are equal for each spin sublevel.)
We note that for $\epsilon=\gamma_{s}/\gamma\lesssim1$
which is the case we expect to apply here due to silicon's indirect
bandgap, the lineshapes are independent of $\epsilon$ and we set
$\epsilon=0.1$. This value should provide an upper bound for the
internal quantum efficiency of our devices. The EQE we estimate is
around 0.05\%, but taking into account the reflection due to the dielectric
constant mismatch at the silicon interface, an upper bound $\epsilon=0.1$
seems reasonable.

The simulations show an enhancement of the EL with magnetic
field due to an increased number of states which overlap with the
singlet ground state. This behavior arises from the fact that the
spin-orbit interaction renormalises the effective hole $g$-factor (with effective Land\'e value $g_{h}/3$ for total spin 3/2 holes),
giving rise to a competition between the exchange interaction and
the effective $g$-factor difference (and hence Zeeman energy)
between electron and hole states. This leads to a mixing between singlet and triplet electron-hole spin configurations with a characteristic saturation field set by the competition between this Zeeman energy difference and the exchange term $B_{sat}\sim J/\Delta g_{eff}\mu_{B}$.
We find that the 300 K experimental lineshape can be reproduced with
$J=-0.75$ meV, which provides an estimate for typical exchange
interactions for transient bound states formed during electron-hole collisions at room temperature in silicon; 
as expected, $J$ is smaller than the estimated exchange energy of 10 meV found for strongly bound excitonic states in Si nanocrystals \cite{brongersma2000size}.
Highlighting the importance of transient bound pairs, we show in the Supplementary Figure 3 that the positive MEL starts to decrease below 150 K, a temperature which matches the exciton binding energy (14.7 meV). This suggests that maximal sensitivity to magnetic field is achieved when the temperature is not too high, allowing interaction effects to show up, but not too small so that electron-hole encounter events do not result in irreversible binding. We emphasise that this model produces a MEL response which does not depend on the direction of the external field, and can therefore explain the positive MEL component in our experiments (Fig. 4). 
While our model reproduces the observed MEL lineshapes, it fails to correctly reproduce the magnitude of the effect. The simulated value in Fig.~5 for $\Delta EL/EL$ at a magnetic field $B$ of 7 Tesla is around $0.1$, as compared to the experimental $0.75$. Theoretical MEL can increase up to $0.76$ in the $\epsilon\gg1$ limit but this would imply a very high internal quantum efficiency, which we do not believe to hold in our devices. Instead, we suggest that multiple recombination attempts, and a more detailed
description of carrier kinetics as well as electron-valley mixing can further amplify the theoretical MEL magnitude. Such a complete theory is beyond the scope of this work.

In conclusion, we have reported a strong increase in the brightness
of silicon LEDs under a magnetic field. These LEDs were fabricated
using a novel technique which allowed us to simultaneously suppress
classical magnetoresistance effects, and obtain effective emission.
In analogy with magneto-optic models developed for organic semiconductors,
we explained our results as arising from the difference in recombination
rates between singlet and triplet electron-hole pairs, allowing the
electron-hole pair exchange energy to be estimated from the experimental
lineshapes. Our investigations suggest an optoelectronic approach to probe spin transport properties in silicon near room temperature, a material with
promise for quantum information processing and spintronics. They also show that the control of spin properties can allow to substantially increase the brightness of SiLEDs which can be important components for chip to chip optical communication. 

{\it Acknowledgments} We thank M. Entin, L. Magaril, A. Rowe and T. Chaneli\`ere for fruitful discussions and acknowledge support from PICS-Royal society, Labex ANR-10-LABX-0039-PALM, and ANR SPINEX. 

{\it Author contributions:} All authors contributed to all aspects of this work.

{\it Methods} 

{\bf Gas immersion Laser doping:} The out-of-equilibrium laser doping was performed in ultra high vacuum (10$^{-9}$ mbar) with a XeCl 308 nm  excimer laser of pulse duration 25 ns and energy 0.96 J/cm$^2$. The precursor gas used for the B(P) doping is ${\rm B Cl_3} (P Cl_3)$. The doping level is finely controlled by the number of laser shots, each shot introducing a fixed B dose corresponding to the surface density of chemisorbtion sites for the precursor gas on the Si surface.  The vertical alignment of the n and p doped spots is performed by illuminating the substrate from the bottom and observing the sample with an infrared camera, thus visualising the bottom spot when doping the top one. 

{\bf Magneto-electroluminescence:} For magneto-electruminescence experiments the samples were mounted inside an optically accessible cryostat magnet providing a static magnetic field up to 7 Tesla. The samples were thermalised at the measurement temperature through a Helium vapor. The electroluminescence was collimated outside the cryostat and focussed on a commercial Silicon or Germanium photodetector a meter away from the cryostat. Taking into account the higher sensitivity of Germanium photodetectors at the silicon emission wavelength the same results were obtained with the two types of photodetectors.

\begin{widetext}

\newpage
\section*{Supplementary Information}
\setcounter{equation}{0}
\renewcommand{\theequation}{S\arabic{equation}}
\setcounter{figure}{0}
\renewcommand\thefigure{S\arabic{figure}}
\renewcommand{\figurename}{Supplementary Figure}

{\bf Supplementary figure \ref{FigTheoS}}

\begin{figure}[h]
\centerline{\includegraphics[clip,width=14cm]{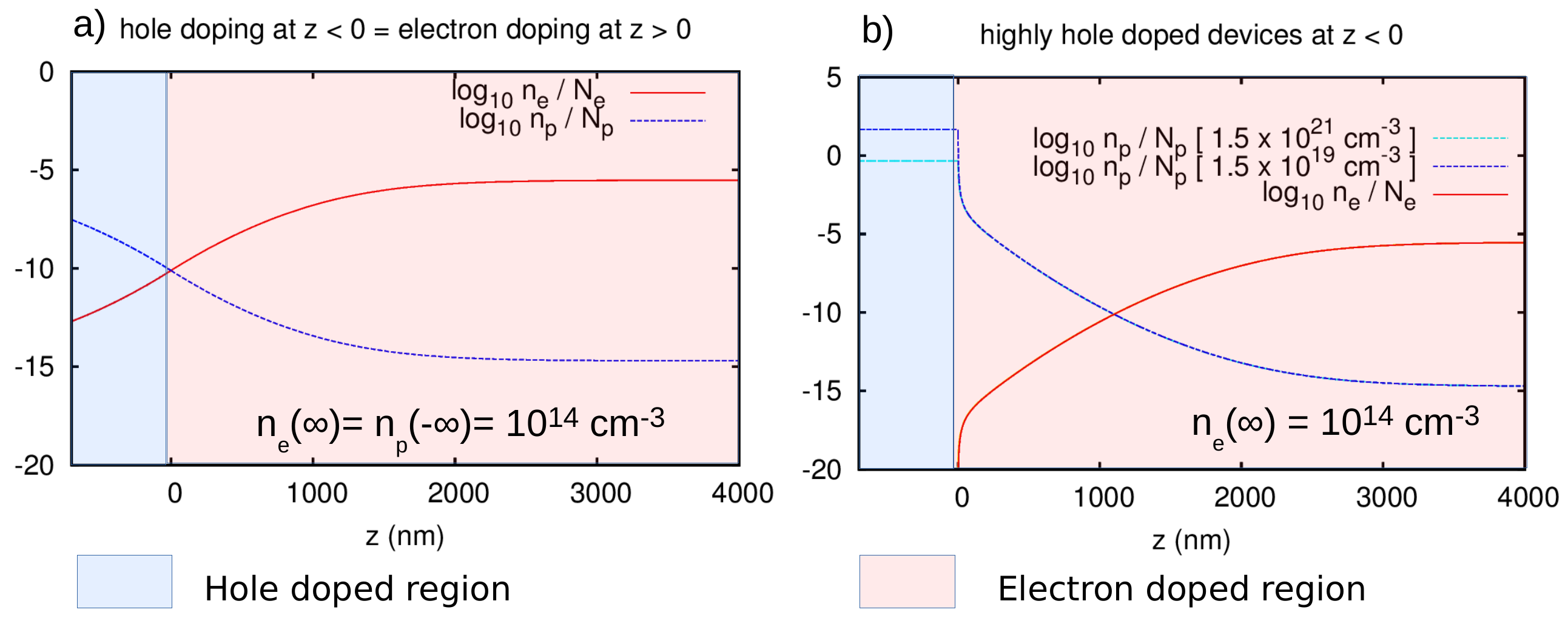}}
\caption{{\bf Theoretical charge density profiles in unbiased devices:} Theoretical dependence of the electron and hole density distributions ($n_e$ and $n_p$ respectively) at the p+/n interface for different doping levels in the $p+$ region as function of the vertical distance to the interface between the p+/n region (the densities are normalized by the effective density of states in the conduction/valence band defined in Eqs.~(\ref{Ne},\ref{Np})) from supplementary note 1 below. The left panel shows the density distribution for equal hole/electron doping in the p+/n regions . This case corresponds to the textbook case of a p/n junction with semiconducting range doping on both sides of the junction. As expected electron and hole densities curves cross in the depletion region of the diode which is centered at the interface at $z = 0$. For high hole doping in the p+ region (right panel) the deletion region is displaced into the n region to within a micron away from the p+/n interface. The electron/hole density profile in the electron doped region then depends only weakly on the doping on the p+ side (almost no change for hole doping densities between $1.5\times10^{19}$ to $1.5 \times 10^{21}\;{\rm cm}^{-3}$). On the other hand the amplitude of the potential barrier (see Figure 2 from main text) increases by 0.5 eV. This confirms our interpretation that the amplitude of the electron blocking potential is the relevant parameter to explain the increase in EL brightness in our experiments.  }
\label{FigTheoS}
\end{figure}

\clearpage 

{\bf Supplementary Figure~\ref{FigELsup}}

\begin{figure}[h]
\centerline{\includegraphics[clip,width=10cm]{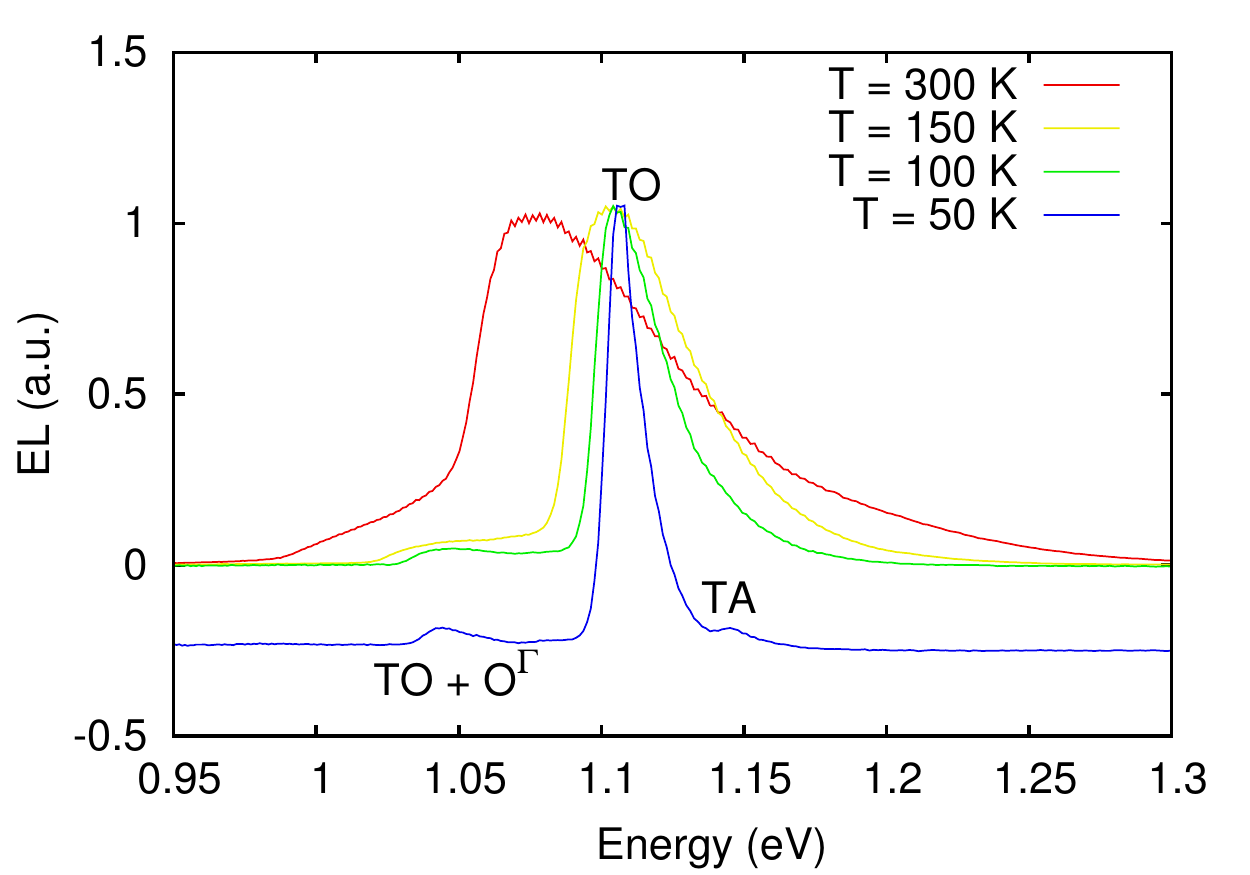}}
\caption{{\bf Low temperature electro-luminescence and EQE measurements:} Normalised electro-luminescence emission spectra from a lateral, $1.2\times10^{21}{\rm cm^{-3}}$ doping device at temperatures from $300$ to $50\;{\rm K}$ at bias current $40mA$. The $50\;{\rm K}$ spectrum, shifted for clarity, reproduces the characteristic low temperature emission spectrum from silicon with well resolved ${\rm TO+O^{\Gamma}}$, TO and TA phonon lines \cite{JangS}. External quantum efficiency (EQE) was measured using a Thorlabs Ge photo-detector (similar results were obtained with a Si photo-detector and a Coherent Ge OP-2 IR detector of 1nW resolution) of known effective area and sensitivity mounted on a goniometer. The emitted power was measured for different inclination angles and as function of the distance between the photodetector and the SiLED confirming the expected scaling of the detected power as function of distance. The quantum efficiency was then deduced by summing contributions from different angles. 
}
\label{FigELsup}
\end{figure}

\clearpage 

{\bf Supplementary Figure~\ref{FigMELtemp} }

\begin{figure}[h]
\centerline{\includegraphics[clip,width=12cm]{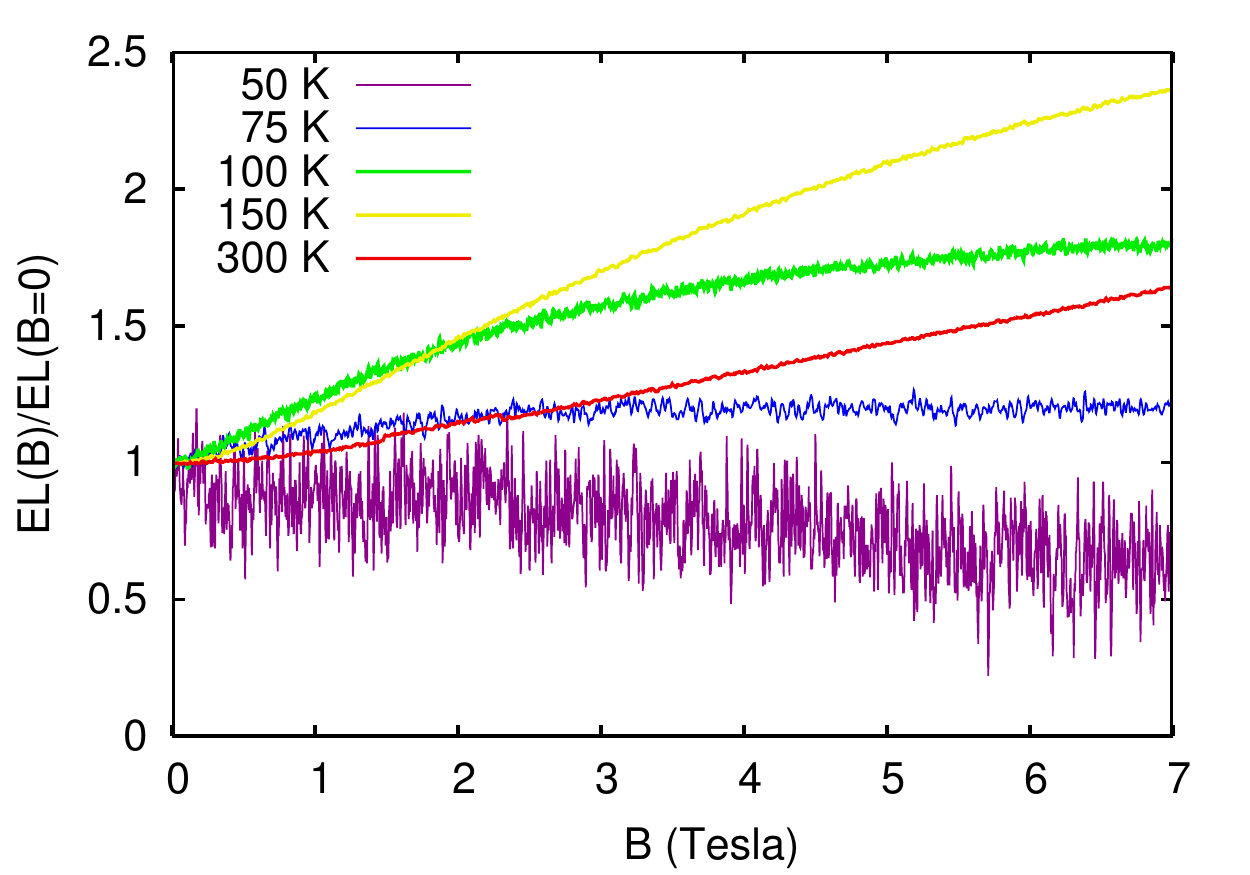}}
\caption{{\bf temperature dependence of the MEL effect:} Temperature dependence of MEL effect in a vertical device with a typical bias current of 20mA. Maximum MEL effect is observed around 150 K. Below this temperature the MEL starts to decrease with a weak residual negative MEL effect at 50 K that we attribute to magneto-diode effects. This characteristic temperatures of 150 K matches the binding energy of excitons in silicon (14.7 meV), suggesting that maximal sensitivity to magnetic field is achieved when the kinetic electron-hole energy is not too high to allow interaction effects to show up, but not to small so that electron-hole encounter events do not result in irreversible binding}.
\label{FigMELtemp} 
\end{figure}

\clearpage

{\bf Supplementary Figure~\ref{FigMagDiode} }

\begin{figure}[h]
\centerline{\includegraphics[clip,width=14cm]{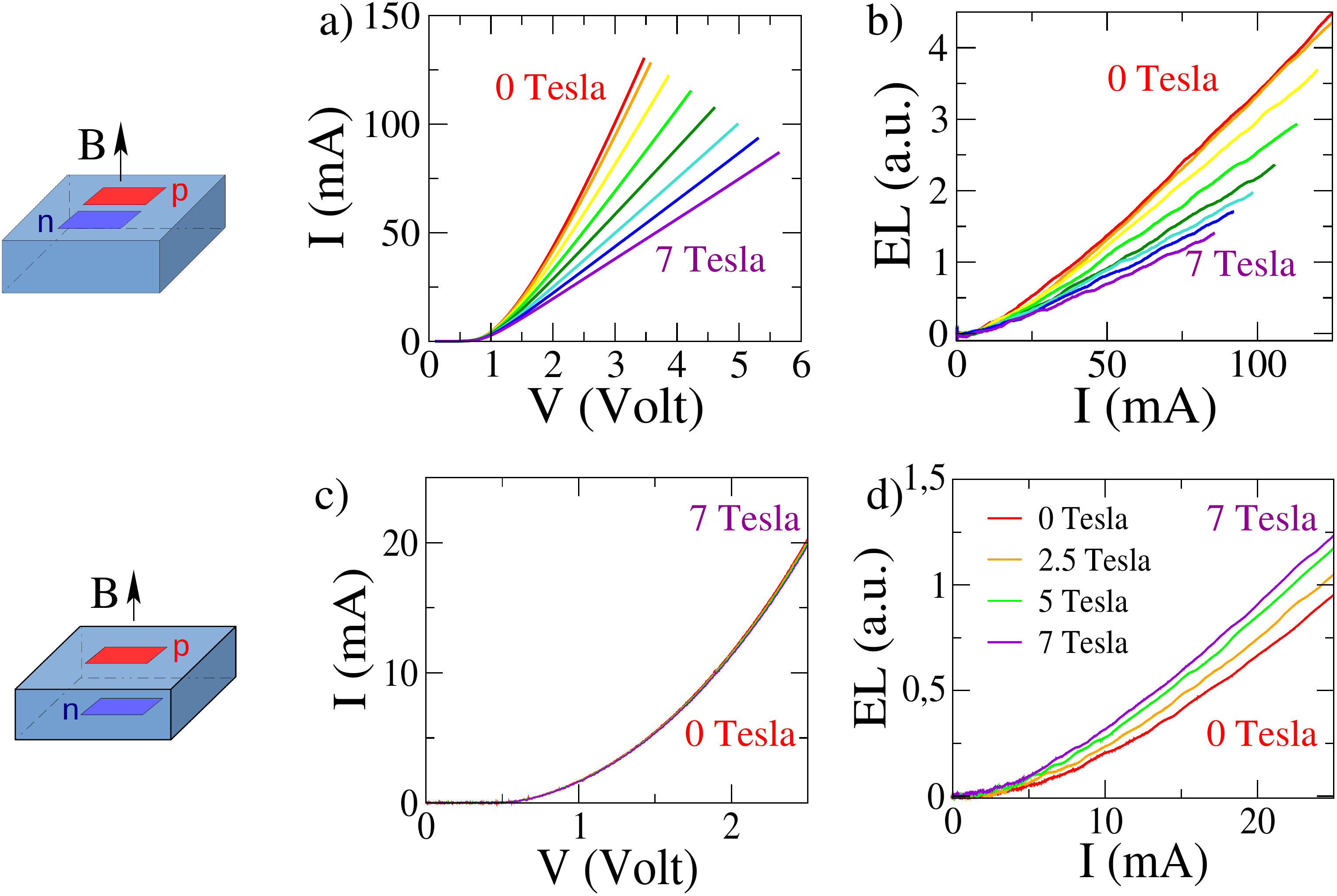}}
\caption{{\bf Comparison of MEL effect in lateral and vertical devices:} Panels a) and b) show the magnetic field effect at room temperature in a lateral devices with doping $1.2\times10^{21}{\rm cm^{-3}}$ where the magnetodiode effect dominates. a) Under magnetic field the resistance of the device increases as electron and hole trajectories are bent by the magnetic field. As the length of the electron-hole trajectory increases the carrier recombination probability is also enhanced, thus a smaller fraction of carriers reaches the interfaces  where radiative recombination is efficient. This leads to a decrease of the EL with magnetic field at fixed current (see panel b). As shown in this figure the MEL is negative for the magnetodiode effect, with MEL and magnetoconductance effects having a similar magnitude. \\
Panels c) and d) show the magnetic field effect at room temperature in vertical devices for perpendicular magnetic field for a vertical device with $1.5\times10^{21}{\rm cm^{-3}}$ doping. As for the $3\times10^{21}{\rm cm^{-3}}$ doping device shown in the main text the magnetoresistance vanishes as the $I(V)$ curve is not changed by magnetic field (panel c) but a substantial increase in EL is observed in panel d) as opposed to the negative MEL in lateral geometry devices (panel b) .
 }
\label{FigMagDiode} 
\end{figure}

\clearpage 

{\bf  Supplementary Figure~\ref{FigELspots} }

\begin{figure}[h]
\begin{tabular}{cc}
\includegraphics[clip,width=14cm]{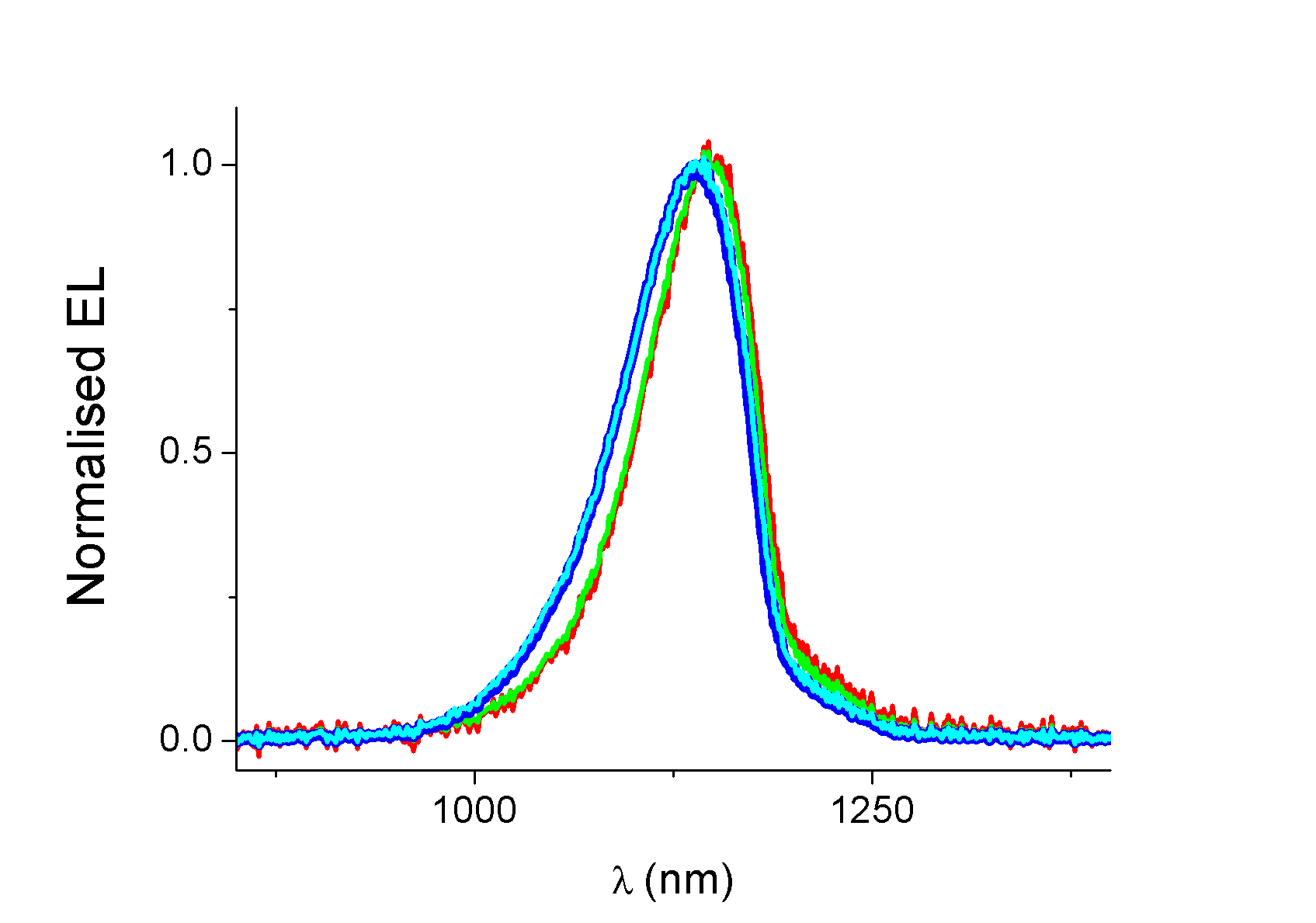}
\end{tabular}
\caption{{\bf Doping dependence of the EL spectra at room temperature:} Spectra (normalized to maximum EL) of lateral (p+/n/n+) SiLED biased with 300 mA, for increasing concentration (red: $1.5\times 10^{20}\rm{ cm^{-3}}$; green: $4.5\times10^{20}{\rm cm^{-3}}$; blue: $1.5\times10^{21}{\rm cm^{-3}}$; light blue: $4.5\times 10^{21}{\rm cm^{-3}}$. At room temperature the EL spectra are almost independent of the doping level. The small difference in line-shape is probably due to slightly different reflection conditions at highly doped interfaces for different doping levels. }
\label{FigELspots} 
\end{figure}

\clearpage 

{\bf  Supplementary Figure~\ref{FigRescaled}}

\begin{figure}[h]
\begin{tabular}{cc}
\includegraphics[clip,width=10cm]{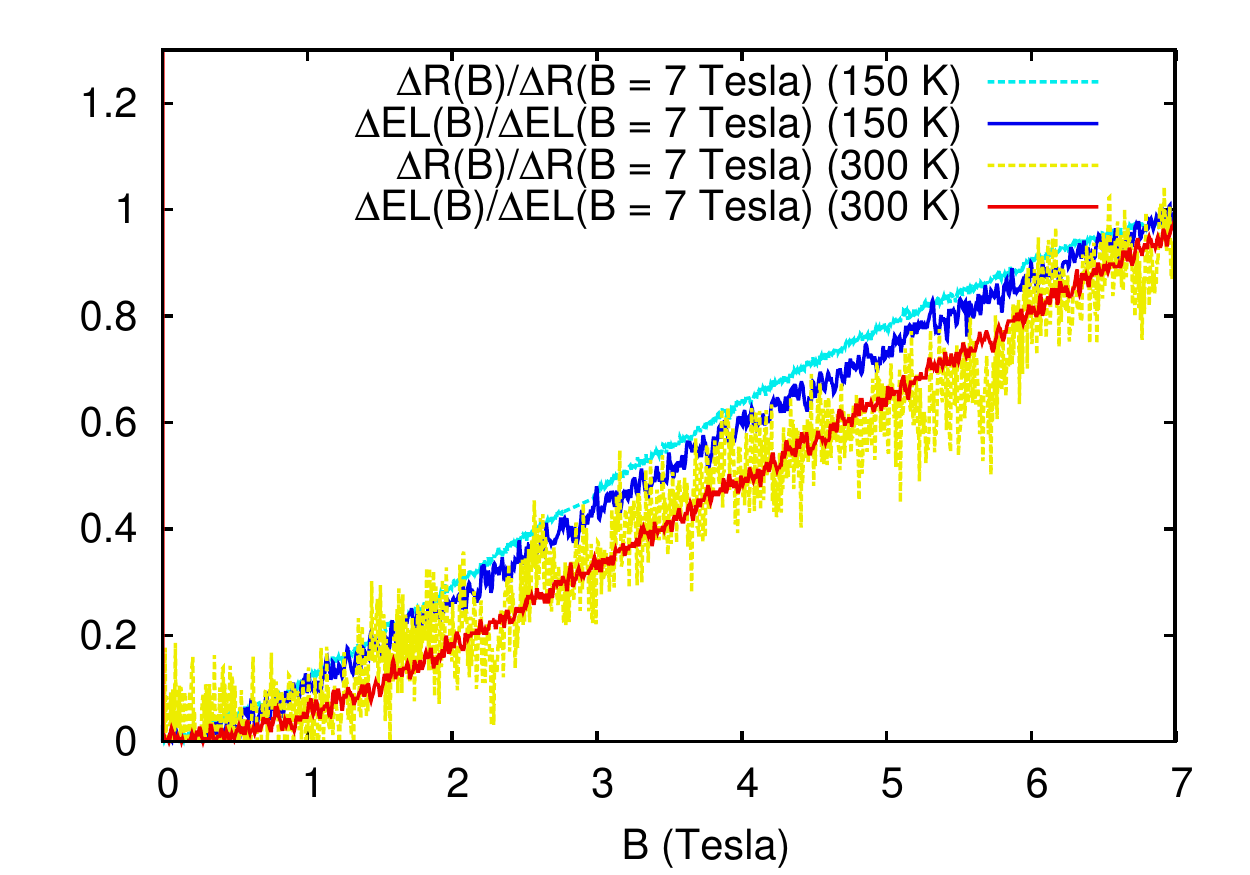}
\end{tabular}
\caption{{\bf Rescaled data for Fig.~3:} Comparison between MEL and MR effects rescaled to their 7 Tesla values for the data shown on Figure 3 from the main text. A small increase in the rescaled resistance consistent with an increasing radiative recombination yield is observed. However the MR effect is much smaller with values at 7 Tesla of around 2.2\% at 150K and 0.5\% at 300K (for comparison MEL is then respectively 275\% and 80\%) and we thus prefer to remain cautious on its interpretation.  }
\label{FigRescaled} 
\end{figure}

\clearpage 

{\bf Supplementary note 1, theoretical charge density profiles in unbiased devices}

To compute the charge distribution in unbiased devices, we solve the Laplace equation on the $V$ the electrostatic potential, $\epsilon_0 \epsilon_r$ the dielectric permittivity, $q$ the charge of the electron:
\begin{align}
\epsilon_r \epsilon_0 \partial_x^2 V = -q [n_e - n_p - C(x)]
\end{align}
where $n_e$ electron density, $n_p$ hole density, $C(x)$ the doping profile.

The densities $n$, $p$ can be determined from the position of the Fermi level $F$ relative to the  bottom of the conduction band $E_c$ and to the top  of the valence band $E_v$ : 
\begin{align}
n_e &= N_e F_{1/2}(\beta F - \beta E_c) \\
n_p &= N_p F_{1/2}(\beta E_v - \beta F)
\end{align}
In the above equation we introduced the notations $\beta = (k_B T)^{-1}$ and $F_{1/2}$ is the Fermi-integral:
\begin{align}
F_{1/2}(y) = \frac{2}{\sqrt{\pi}} \int_0^{\infty} \frac{\sqrt{t} dt}{1 + e^{t - y}} 
\end{align}
and $N_e$, $N_P$ which are effective density of states in the conduction/valence bands. They are given by the following formulas \cite{Zhe}
\begin{align}
N_e &= 12 \left( \frac{m_e^* k_B T}{2 \pi \hbar^2} \right)^{3/2} \label{Ne} \\
N_p &= 2 \left( \frac{m_p^* k_B T}{2 \pi \hbar^2} \right)^{3/2} \label{Np}
\end{align}
which lead to $N_e = 2.8\times10^{19}\;{\rm cm^{-3}}$ and $N_p = 1\times10^{19}\;{\rm cm^{-3}}$ at room temperature.

Finally the electro-chemical potential $\mu$ remains constant across the bilayer :
\begin{align}
\mu = F + q V 
\end{align}

This leads to the following self-consistent equation on the electrostatic potential (we now count the potential from the electro-chemical potential $\mu$, thus $F = - q V$):
\begin{align}
\epsilon_r \epsilon_0 \partial_x^2 V = -q [N_e F_{1/2}(-q V \beta - E_c \beta) - N_p(q V \beta + E_v \beta) - C(x)]
\end{align}

Boundary conditions on the electrostatic potential are given by:
\begin{align}
N_e F_{1/2}(-q V(\pm \infty) \beta - E_c \beta) - N_p(q V(\pm \infty) \beta + E_v) = C(\pm\infty)
\end{align}

Choosing dimensionless units: 
\begin{align}
\partial_x^2 V = -F_{1/2}(-V - \beta \Delta/2) + N_r F_{1/2}(V - \beta \Delta/2) + C(x)
\end{align}
with $N_r = N_p/N_e$.

Here the potential is in units of temperature $k_B T = \beta^{-1}$, $C(x)$ in units of $N_e$ and the length-scale $\lambda$ is set by
\begin{align}
\lambda = \sqrt{\frac{\epsilon_R \epsilon_0}{q \beta N_e}}
\end{align}
with room temperature values for silicon: 
\begin{align}
\beta^{-1} &= 26\;{\rm meV}\\
\lambda &= 0.72\;{\rm nm}\\
\beta \Delta &= 43 
\end{align}

Thus we are thus finally lead to the equation: 
\begin{align}
\partial_x^2 V &= -F_{1/2}(-V - \beta \Delta/2) + N_r F_{1/2}(V - \beta \Delta/2) - C_p \eta(-x) + C_e \eta(x)\\
 &= \partial_V \left[ F_{3/2}(-V - \beta \Delta/2) + N_r F_{3/2}(V - \beta \Delta/2) \right] - C_p \eta(-x) + C_e \eta(x)
\end{align}
where all the coefficients are known ($F_{3/2}(x)$ is the complete Fermi-integral).

For $x > 0$, this can be integrated to:
\begin{align}
H_e &= \frac{(\partial_x V)^2}{2} - \left[ F_{3/2}(-V - \beta \Delta/2) + N_r F_{3/2}(V - \beta \Delta/2) + C_e V \right] 
\end{align}
and for $x < 0$
\begin{align}
H_p = \frac{(\partial_x V)^2}{2} - \left[ F_{3/2}(-V - \beta \Delta/2) + N_r F_{3/2}(V - \beta \Delta/2) - C_p V \right] 
\end{align}

Combining the two conservation laws allows us to find the potential $V(0)$:
\begin{align}
H_p-H_e &= (C_e+C_p) V(0) 
\end{align}
Starting from $V(0)$ the potential and density profiles can then be obtained by direct integration of the equations of motion.

For devices under bias the drift-diffusion equations were solved using a finite elements method.

{\bf Supplementary references :} 


\bibitem{Zhe} S.M. Sze, {\it Physics of semiconductor devices}, A Wiley, ISBN 0-471-09837-X

\bibitem{JangS} Miin-Jang Chen, Eih-Zhe Liang, Shu-Wei Chang, and Ching-Fuh Lin a), Jour. of Appl. Phys. {\bf 90} 789 (2001)


\end{widetext}

\end{document}